\documentclass{bmvc2k}
\usepackage{multirow}
\usepackage{amsfonts}
\usepackage{wrapfig}
\DeclareMathOperator*{\argmin}{min}

\usepackage{pifont}
\newcommand{\cmark}{\ding{51}}
\newcommand{\xmark}{\ding{55}}

\title{SAGAN: Adversarial Spatial-asymmetric Attention for Noisy Nona-Bayer Reconstruction}

\addauthor{S M A Sharif}{sma.sharif.cse@ulab.edu.bd}{1}
\addauthor{Rizwan Ali Naqvi}{rizwanali@sejong.ac.kr}{2}
\addauthor{Mithun Biswas}{mithun.bishwash.cse@ulab.edu.bd}{1}

\addinstitution{
 Rigel-IT, Bangladesh
}
\addinstitution{
 Sejong University, South Korea
}

\runninghead{SAGAN}{Preprint}

\begin{document}

\maketitle
\vspace{-.4cm}
\begin{abstract}
Nona-Bayer colour filter array (CFA) pattern is considered one of the most viable alternatives to traditional Bayer patterns. Despite the substantial advantages, such non-Bayer CFA patterns are susceptible to produce visual artefacts while reconstructing RGB images from noisy sensor data. This study addresses the challenges of learning RGB image reconstruction from noisy Nona-Bayer CFA comprehensively.  We propose a novel spatial-asymmetric attention module to jointly learn bi-direction transformation and large-kernel global attention to reduce the visual artefacts. We combine our proposed module with adversarial learning to produce plausible images from Nona-Bayer CFA. The feasibility of the proposed method has been verified and compared with the state-of-the-art image reconstruction method. The experiments reveal that the proposed method can reconstruct RGB images from noisy Nona-Bayer CFA without producing any visually disturbing artefacts. Also, it can outperform the state-of-the-art image reconstruction method in both qualitative and quantitative comparison. Code available: \url{https://github.com/sharif-apu/SAGAN_BMVC21}.
\end{abstract}

\vspace{-.4cm}
\section{Introduction}

The past decade has experienced a revolutionary takeover in mobile photography. Explicitly, the alleviation in computation photography and innovation on mobile hardware allows the original equipment manufacturers (OEMs) to provide handy experiences to the mobile photographers. However, the perceptual quality of smartphone cameras still incorporates notable drawbacks due to the smaller sensor size and unable to deliver professional-grade image quality in stochastic lighting conditions \cite{a2021beyond,ignatov2020replacing}. Contrarily, enlarging the sensor size of the mobile cameras always remains a strenuous process. Explicitly, the compact nature of mobile devices holds back the OEMs to perceive a substantial push in the sensor size.  To address such an inevitable dilemma, many OEMs have leverage pixel enlarging techniques known as pixel-binning with a non-Bayer CFA pattern \cite{a2021beyond, kim2019deep,lahav2010color, barna2013method}.  Among such non-Bayer CFA patterns, Nona-Bayer has illustrated widespread practicability over its Bayer counterparts.

\begin{wrapfigure}{!htb}{0.4\textwidth}
\begin{center}
\begin{tabular}{cc}
\bmvaHangBox{{\includegraphics[width=2.2cm]{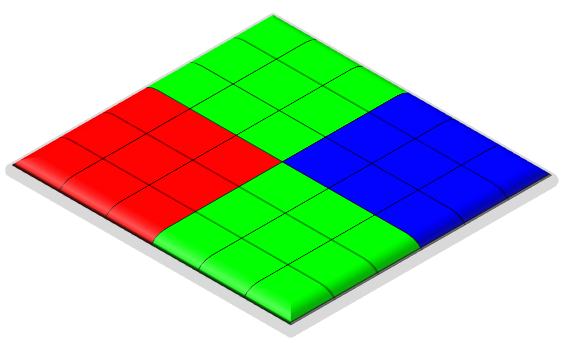}}}&
\bmvaHangBox{{\includegraphics[width=2.2cm]{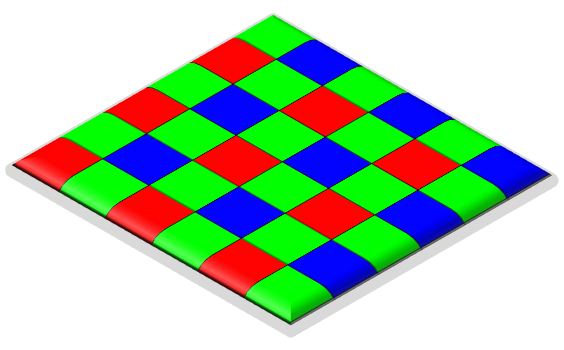}}}\\
(a)&(b)
\end{tabular}
\caption{ Comparison between CFA patterns. (a) Nona-Bayer CFA. (b) Bayer CFA. }
\label{fig:cfa}
\end{center}
\vspace{-.5cm}
\end{wrapfigure}

Typically, a Nona-Bayer CFA pattern comprises of three consecutive homogenous pixels in the vertical and horizontal direction, as shown in Fig. \ref{fig:cfa}. Notably, such a CFA pattern allows the sensing hardware to combine homogenous pixels into a bigger pixel to gather up to three times higher light intensity in stochastic lighting conditions. Apart from the improving low-light performance, a Nona-Bayer CFA concedes the practicability of higher resolution sensors in mobile devices and allows to produce high definition contents (i.e., 8K videos) with a natural bokeh effect.  Hence, most recent flagship smartphones like Samsung S20 Ultra, Note 20 Ultra, S21 Ultra, Xiaomi Mi 11 Ultra, etc., have utilized such Nona-Bayer CFA on top of the 108-megapixel image sensor to deliver a versatile photography experience to enthusiastic mobile photographers.

\begin{figure}[!htb]
\begin{center}
\begin{tabular}{cccc}

\multirow{2}{*}{\bmvaHangBox{{\includegraphics[width=3.0cm, height=2.81cm]{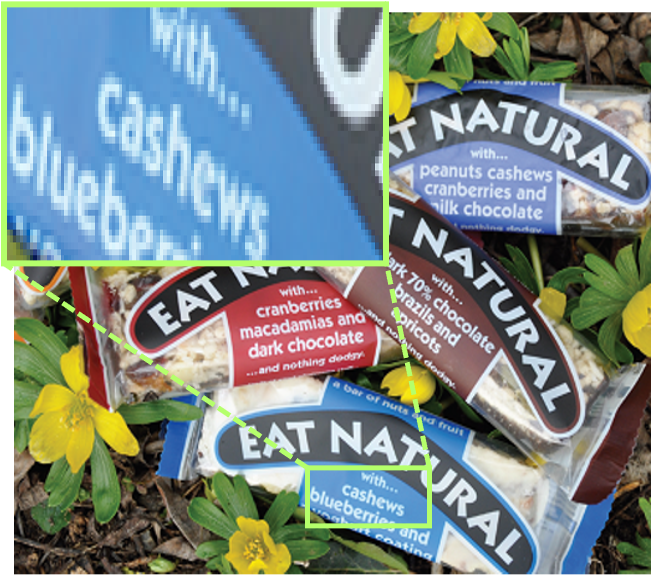}}}}  & 

\hspace{-.45cm}
\bmvaHangBox{{\includegraphics[width=2.0cm, height=1.13cm]{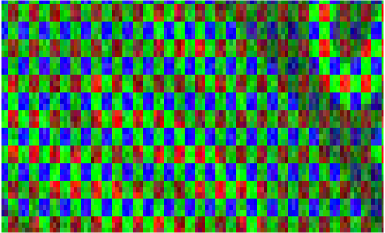}}} & 
\hspace{-.45cm}
\bmvaHangBox{{\includegraphics[width=2.0cm, height=1.13cm]{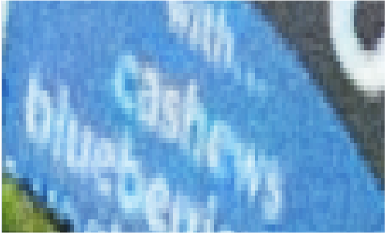}}}& 
\hspace{-.45cm}
\bmvaHangBox{{\includegraphics[width=2.0cm, height=1.13cm]{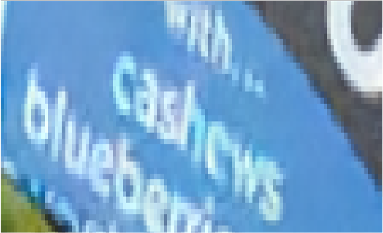}}}\\
&(b)&(c)&(d)\\

& \hspace{-.45cm} \bmvaHangBox{{\includegraphics[width=2.0cm, height=1.13cm]{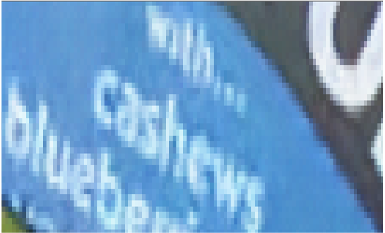}}}& \hspace{-.45cm}
\bmvaHangBox{{\includegraphics[width=2.0cm, height=1.13cm]{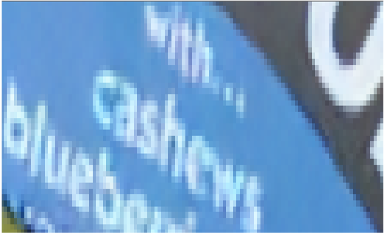}}}& \hspace{-.45cm}
\bmvaHangBox{{\includegraphics[width=2.0cm, height=1.13cm]{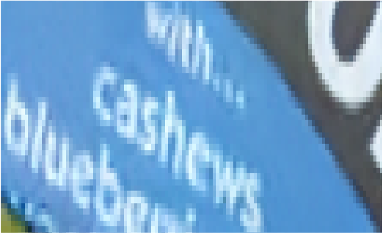}}}\\
(a)&(e)&(f)&(g)

\end{tabular}
\caption{ Noisy Nona-Bayer reconstruction with state-of-the-art image reconstruction methods at $\sigma=30$. (a) Ground-truth RGB Image. (b) Noisy Nona-Bayer Input. (c) Deepjoint \cite{gharbi2016deep}. (d) Kokkinos \cite{kokkinos2018deep}. (e) DPN \cite{kim2019deep}. (f) BJDD \cite{a2021beyond}. (g) \textbf{SAGAN (Ours)}}
\label{fig:introComp}
\end{center}
\vspace{-.5cm}
\end{figure}

Despite numerous advantages, reconstructing an RGB image from a Nona-Bayer CFA is a challenging task. It is worth noting, the distance of homogenous pixels between two recurring Nona-Bayer CFA patterns are three-time larger than a typical Bayer CFA (please see Fig. \ref{fig:cfa}. Subsequently, any complex composition like text with a distinct background that appears between two consecutive patterns can produce visual artefacts. Moreover, the substantial sensor noise along with artefact-prone CFA pattern makes the reconstruction process notably complicated \cite{a2021beyond}. We found even the state-of-the-art deep image reconstruction method (i.e., joint demosaicing and denoising (JDD), non-Bayer reconstruction methods) illustrates notable shortcomings in reconstructing RGB images from a noise-contaminated Nona-Bayer CFA pattern. In most instances, the existing methods tend to produce structural distortion and false colour artefacts, as shown in Fig. \ref{fig:introComp}.

To address the deficiencies of existing works, we propose a novel learning-based JDD method for Nona-Bayer reconstruction. To the best concern, this is the first work in the open literature that introduces an end-to-end deep model for reconstructing RGB images from a noisy Nona-Bayer CFA pattern. Our proposed method incorporates a novel spatial-asymmetric attention module to reduce visual artefacts from reconstructed RGB images. Our proposed module learns attention over the vertical and horizontal transformation of a Nona-Bayer CFA and combines it with large-kernel global attentions. Additionally, we proposed an adversarial (a.k.a generative adversarial network (GAN) \cite{goodfellow2014generative}) guidance with our spatial-asymmetric attention for producing visually plausible images. We denoted our proposed method as spatial-asymmetric attention GAN (SAGAN) in the rest of the paper. The practicability of the proposed method has been extensively studied with the benchmark dataset and compared with state-of-the-art deep reconstruction methods. The major contributions of the proposed method have summarized as follows: 1) Proposes and illustrates the practicability of an end-to-end deep network for performing image reconstruction from challenging noisy Nona-Bayer CFA pattern images. 2) Proposes a novel spatial-asymmetric attention module to reduce visual artefacts and combined it with adversarial training to produce plausible images. 3) Compare and outperform existing learning-based reconstruction method in both qualitative and quantitative comparison. 

\vspace{-.5cm}
\section{Related Works}
The related works of our proposed method have briefly described in this section.

\textbf{Joint demosacing and denoising.}
Noise suppression with reconstructing RGB images from CFA patterns have illustrated a significant momentum in recent years. In practice, such JDD manoeuvres can significantly improve the perceptual quality of final reconstructed images. In the early days, JDD was mostly performed with optimization-based strategies  \cite{hirakawa2006joint, tan2017joint}.  However, in recent time, deep learning has takeover the limelight from its traditional counterparts by learning JDD from a convex set of data samples. 

In recent work, \cite{gharbi2016deep} trained an end-to-end deep network to achieve state-of-the-art performance in Bayer JDD. Later, \cite{kokkinos2018deep} combined deep residual denoising with a majorization-minimization technique to perform JDD on the same CFA pattern. Similarly, \cite{liu2020joint} also proposed a deep method with density-map and green channel guidance to outperform their previous JDD methods.  Apart from the Bayer JDD, a recent study \cite{a2021beyond} proposed a deep network to perform JDD on Quad Bayer CFA. Notably, \cite{a2021beyond} has illustrated that visual attention with perceptual optimization can significantly accelerate the performance of non-Bayer JDD.

\textbf{Non-Bayer Reconstruction.} Quad Bayer CFA shared similar characteristics as a Nona-Bayer CFA and widely used in recent smartphones cameras. A recent study \cite{kim2019deep} has proposed a duplex pyramid network for reconstructing the Quad Bayer CFA pattern. Similarly, \cite{kim2021under} proposed to learn an under-display camera pipeline exclusively for Quad Bayer CFA.

\textbf{Attention Mechanism.} The concept of attention mechanisms intends to focus on the important features as similar to the human visual system.  In the past decade, many works have incorporated novel attention mechanisms for accelerating different vision tasks. In recent work,\cite{hu2018squeeze} proposed a squeeze-and-excitation network for achieving channel-wise attention for accelerating image classification. \cite{wang2017residual} proposed a residual attention network for having 3D attention over intermediate features. Later, \cite{woo2018cbam} proposed a lightweight convolutional block attention module to accelerate the learning process of feed-forward networks. Similarly, \cite{yu2019free} proposed a convolutional attention mechanism to learn dynamic feature attentions. It is worth noting, none of the existing methods has exploited the visual attention on asymmetric manner. In this study, we depicted that such spatial-asymmetric attention can significantly improve the performance of low-level vision task, explicitly the Nona-Bayer reconstruction. 

\vspace{-.4cm}
\section{Method}
\label{sec:method}
This section describes the proposed method as well as our SAGAN architecture.

\begin{figure}[!htb]
\begin{center}
\bmvaHangBox{{\includegraphics[width=.95\textwidth]{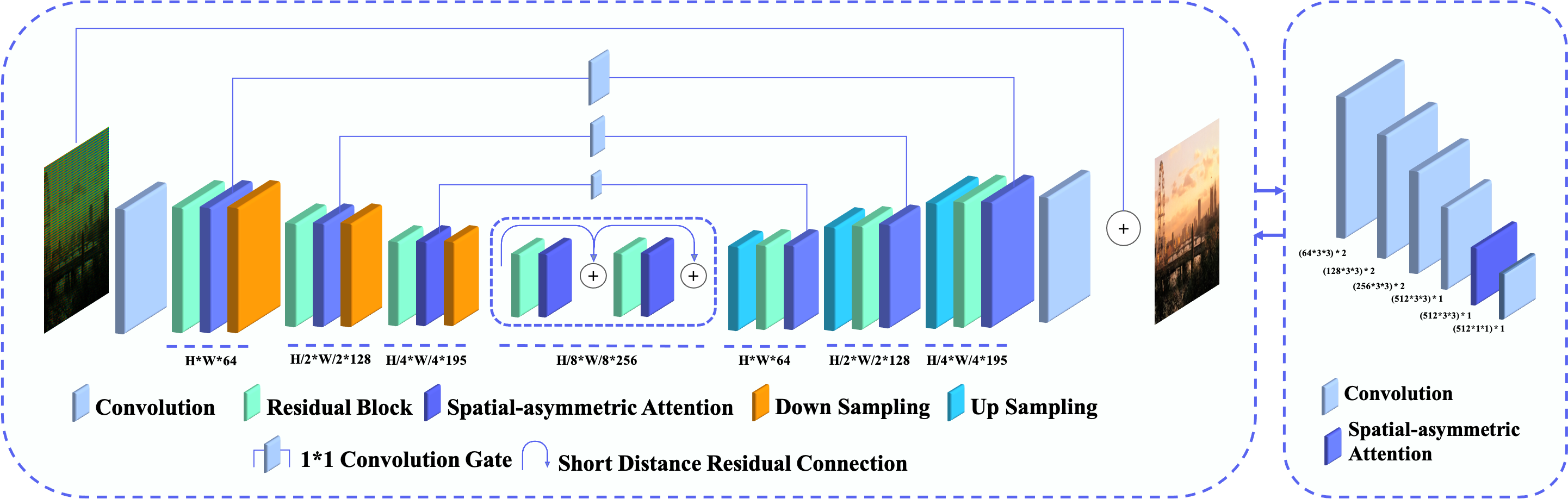}}}\\
\caption{ Overview of the proposed method. Our SAGAN comprises a novel spatial-asymmetric module and guided by adversarial training.}
\label{fig:overview}
\end{center}
\vspace{-.5cm}
\end{figure}


\subsection{Network Design}

Fig. \ref{fig:overview} illustrates the overview of the proposed SAGAN architecture. The proposed method has been designed as a deep network incorporating novel spatial-asymmetric attention along with adversarial training. Our generative method ($\mathrm{S}$) learns to translate a Nona-Bayer mosaic pattern ($\mathbf{I_N}$) as $\mathrm{S}: \mathbf{I_N} \to  \mathbf{I_R}$. Here, ($\mathbf{I_R}$) present the reconstructed RGB image as $\mathbf{I_R} \in [0,1]^{H \times W \times 3}$. $H$ and $W$ represent the height and width of the input mosaic patterns and output RGB images. 

\begin{wrapfigure}{!htb}{0.45\textwidth}
\vspace{-1.5cm}
\begin{center}

\bmvaHangBox{{\includegraphics[width=0.43\textwidth]{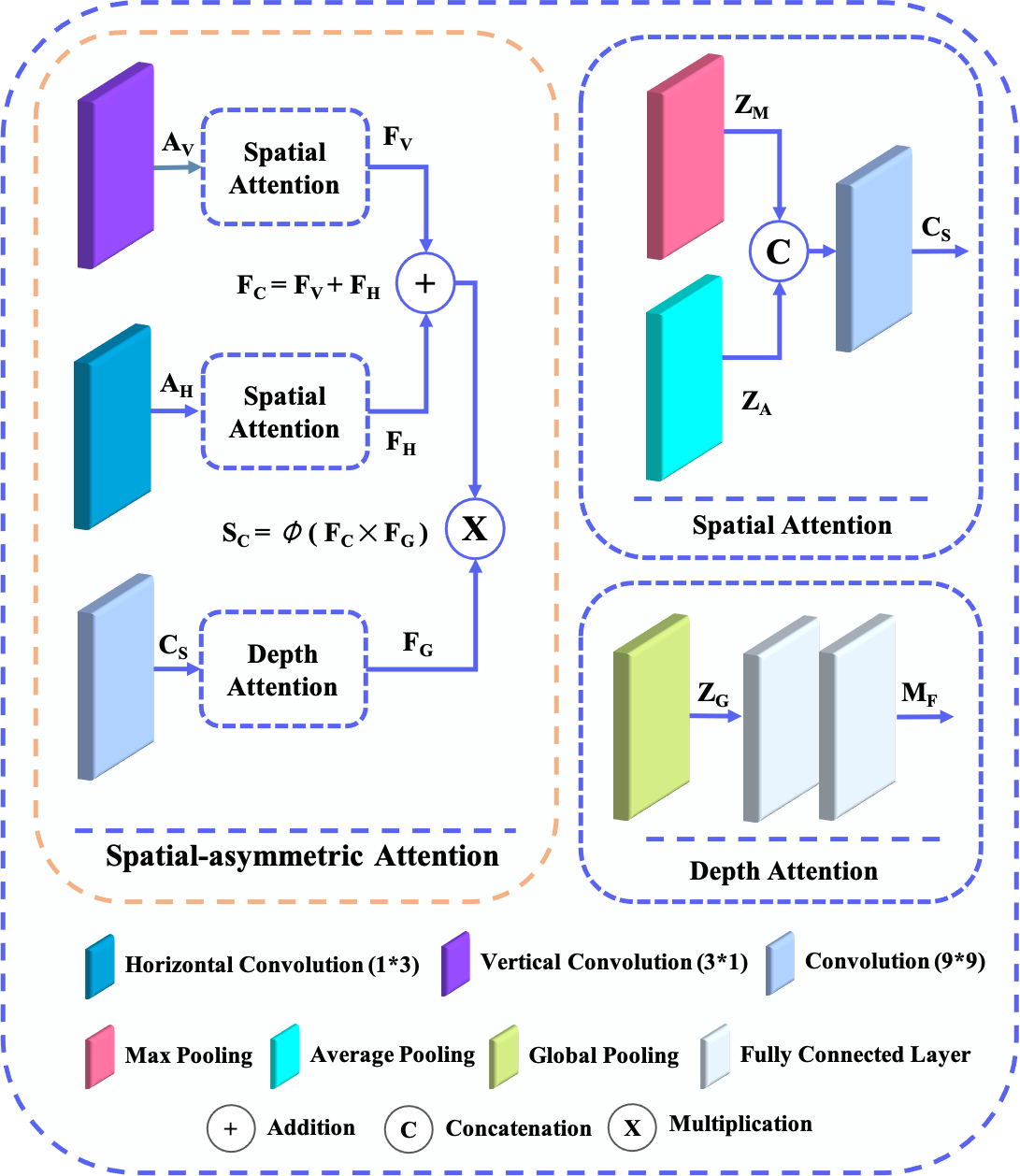}}}\\
\caption{ Overview of proposed spatial-asymmetric attention module. Our proposed block aims to substantially reduce visual artefacts, which typically arises due to Nona-Bayer CFA.}
\label{fig:stab}
\end{center}
\end{wrapfigure}

\subsubsection{Spatial-asymmetric Attention Module}
The proposed spatial-asymmetric attention module intends to reduce visual artefacts from the reconstructed RGB images. As Fig. \ref{fig:stab} depicts, we leverage asymmetric convolution operations \cite{ding2019acnet,lo2019efficient} to extract a sequence of vertical and horizontal feature maps. Later, we utilized spatial attention \cite{woo2018cbam} over the extracted horizontal and vertical features to perceive a pixel-level feature suppression/expansion as follows:

\begin{equation}
\mathbf{F_V} = \tau(\mathbf{C_S}([\mathbf{Z_{A}(\mathbf{A_V}(X))};\mathbf{Z_M(\mathbf{A_V}(X))}])) 
\end{equation}

\begin{equation}
\mathbf{F_H} = \tau(\mathbf{C_S}([\mathbf{Z_{A}(\mathbf{A_H}(X))};\mathbf{Z_M(\mathbf{A_H}(X))}])) 
\end{equation}

Here, $\mathbf{A}(\cdot)$, $\mathbf{C}(\cdot)$, and $\tau$ represent the asymmetric convolution operation, square convolution  and sigmoid activation, respectively. Additionally, $\mathbf{Z_{A}}$ and $\mathbf{Z_{M}}$ present the average pooling and max pooling, which generates two 2D feature maps as $\mathbf{X_{A}} \in \mathbb{R}^{1 \times H \times W}$, $\mathbf{X_{M}} \in \mathbb{R}^{1 \times H \times W}$ and concatenated into a single 2D feature map.

An aggregated bi-directional attention over a given feature map has obtained as:

\begin{equation}
\mathbf{F_C} =  \mathbf{F_V} + \mathbf{F_H}
\end{equation}

Apart from the asymmetric attention, we also appropriated a squeeze-extractor descriptor \cite{hu2018squeeze} to learn depth-wise attention on a globally extracted feature map as follows:

\begin{equation}
\mathbf{F_G} = \mathbf{M_F(\mathbf{Z_{G}(\mathbf{C_S}(X)})}) 
\end{equation}

Here, $\mathbf{M_F}$ and $\mathbf{Z_{G}}$ present consecutive fully connected layers and global pooling operations.

Notably, our spatial-asymmetric module incorporated large-kernel convolution (i.e., $9 \times 9$) operations. Here, these square convolution operations are intended to learn global image correction by exploiting larger reception fields \cite{peng2017large}. We obtained the final output of the spatial-asymmetric attention module as follows:

\begin{equation}
\mathbf{S_A} = \phi(\mathbf{F_C} \times  \mathbf{F_G}) 
\end{equation}

Here, $\phi$ denotes the leaky ReLU activation function.

\subsubsection{SAGAN Generator}

The proposed SAGAN generator has been designed as well-known U-Net architecture with convolutional features gates \cite{sharif2020learning,jiang2021enlightengan,a2021two}. Our SAGAN generator utilized multiple feature depth levels (i.e., 64, 128, 192, and 256) for feature encoding-decoding. Each feature level of the proposed generator comprises a residual block and a spatial-asymmetric attention block. Here, the residual blocks are intended to accelerating denoising performance, while spatial-asymmetric attention blocks are intended to reduce visual artefacts.  We obtained downsampling and upsampling using square convolutional (with stride = 2) and pixel-shuffles upsampling operations. Apart from that, our SAGAN generator also comprises two consecutive middle blocks with a short distance residual connection. Additionally, it connects multiple layers of encoder-decoder with $1 \times 1$ convolutional feature gates. Here, the short distance residual connection and the convolutional gates help our SAGAN converging with informative features.

\subsubsection{SAGAN Discriminator}
The architecture of our SAGAN discriminator has been designed as a stacked convolutional neural network (CNN). The first seven layers of the proposed discriminator are $3 \times 3$ convolution layers, which we normalized with batch normalization and activated with a swish activation. The convolutional layers are followed by a spatial-asymmetric attention module and $1 \times 1$ convolutional output layer with sigmoid activation. Here, every $(2n-1)^{th}$ layer of our SAGAN discriminator reduces its spatial dimension by incorporating a stride = 2.

\subsection{Optimization}

The proposed SAGAN has been optimized with a multi-term objective function. For a given  training set $\{ \mathbf{I_N}^t, \mathbf{I_G}^t \}_{t=1}^P$ consisting of $N$ image pairs, the training process aims to minimize the objective function describes as follows: 
 
\begin{equation}
 W^\ast = \arg{\argmin_W}\frac{1}{P}\sum_{t=1}^{P}\mathcal{L}_{\mathit{S}}(\mathrm{S}(\mathbf{I_N}^t), \mathbf{I_G}^t)
 \label{fLoss}
\end{equation}

Here, $\mathcal{L}_{\mathit{S}}$ represent the proposed SAGAN loss, and $W$ presents the parameterised weights of the SAGAN generator.

\textbf{Reconstruction Loss.} Our proposed SAGAN loss comprises an L1-norm as standard reconstruction loss as follows:

\begin{equation}
 \mathcal{L}_{\mathit{R}} = \parallel \mathbf{I_G}-\mathbf{I_R} \parallel_1
\end{equation}

Here, $\mathbf{I_R}$ presents the reconstructed RGB output of $\mathrm{S}$ and $\mathbf{I_G}$ presents the ground-truth RGB image.

\textbf{Perceptual Colour Loss (PCL).} Apart from the reconstruction loss, we leverage a perceptual colour loss \cite{a2021beyond} to perceive a consistent colour accuracy across different colour spaces. Here, the perceptual colour loss is obtained as follows:

\begin{equation}
 \mathcal{L}_{\mathit{C}} = \Delta{E} \Big( \mathbf{I_{G}}, \mathbf{I_R} \Big)
\end{equation}

Here, $\Delta{E}$ represents the CIEDE2000 colour difference \cite{luo2001development}, which has calculated by comparing reconstructed image ($\mathbf{I_R}$) and the ground-truth image ($\mathbf{I_G}$).

\textbf{Adversarial Loss.} The proposed SAGAN leverages adversarial training to produce natural colour while retaining texture information. Therefore, discriminator maximise a loss $\mathbb{E}_{X, Y}$ as: $\mathbb{E}_{X, Y} \big[\log D\big(X, Y\big) \big]$. Contrarily, our SAGAN generator aims to minimize the generator loss as follows:

\begin{equation}
 \mathcal{L}_{\mathit{G}}= - \sum_{t} \log D(\mathbf{I_R}, \mathbf{I_G})
\end{equation}

\textbf{SAGAN loss.} We perceived SAGAN loss by adding individual losses as follows:

\begin{equation}
 \mathcal{L}_{\mathit{S}}= \mathcal{L}_{\mathit{R}} + \mathcal{L}_{\mathit{C}} + \lambda_{G}.\mathcal{L}_{\mathit{G}}
\end{equation}

Here, $\lambda_{G}$ presents the adversarial regulators, which has been tuned arbitrarily as $\lambda_{G} = 1e-4$ for stabilizing our adversarial training.

\section{Experiments and Results}
The practicability of the proposed SAGAN has verified with dense experiments. This section details the experiments and discusses the results.

\subsection{Experiment Setup}
We extracted a total of 741,968 non-overlapping  $128 \times 128$ image patches from DIV2K \cite{agustsson2017ntire}, Flickr2K \cite{timofte2017ntire}, HDR+ \cite{hasinoff2016burst} datasets to learn noisy image reconstruction. We presumed that JDD performed after non-linear mapping and was independent of additional ISP tasks similar to previous works \cite{a2021beyond, kokkinos2018deep, gharbi2016deep}. Subsequently, we sampled sRGB images according to the CFA pattern and contaminated the sampled images with random noise as ($\mathcal{N}\mathbf{I_N}|\sigma)$. Here, $\sigma$ represents the standard deviation of the random noise distribution, which is generated by $\mathcal{N}(\cdot)$ over a sampled input $\mathbf{I_N}$. We evaluated our method in both sRGB and Linear RGB colour spaces. To evaluate our method in sRGB space, we combined multiple sRGB benchmark dataset including BSD100 \cite{MartinFTM01}, McM \cite{wu2011single}, Urban100 \cite{cordts2016cityscapes}, Kodak \cite{yanagawa2008kodak}, WED \cite{ma2016waterloo} into an unified dataset. Apart from that, we included linear RGB images from MSR demosaicing dataset \cite{khashabi2014joint}, which has been denoted as Linear RGB in later sections.

Apart from learning from synthesized dataset, we studied our proposed method with real-world noisy data samples also. Therefore, we trained our SAGAN with real-world noisy sampled images from Smartphone Image Denoising Dataset (SIDD) \cite{abdelhamed2018high, abdelhamed2019ntire}. Also, we developed an android application to capture noisy images with real Nona-Bayer hardware.  Later, We incorporated a Samsung Galaxy Note 20 Ultra hardware (i.e., 108MP Nona-Bayer sensor) to collect Nona-Bayer captures for evaluating our SAGAN on real-world scenarios.

We implemented our SAGAN in the PyTorch \cite{pytorch} framework. The generator and discriminator of the proposed network have optimized with an Adam optimizer \cite{kingma2014adam}, where hyperparameters are set as $\beta_1 = 0.9$, $\beta_2 = 0.99$, and learning rate = $5e-4$. We trained our generator and discriminator jointly for 200,000 steps with a constant batch size of 16. It took around 120 hours to converge with given data samples.  We employed a graphical Nvidia Geforce GTX 1060 (6GB) graphical processing unit (GPU) to conduct our experiments.

\subsection{Comparison with State-of-the-art Methods}

The performance of SAGAN has been studied with different CFA Patterns (i.e., Nona-Bayer and Bayer CFA patterns) and compared with state-of-the-art reconstruction methods. We included deep Bayer joint demosaicking and denoising methods like Deepjoint  \cite{gharbi2016deep}, Kokkinos \cite{kokkinos2018deep}, Non-Bayer JDD method like BJDD \cite{a2021beyond}, and Quad Bayer reconstruction method like DPN \cite{kim2019deep} for the comparison. For a fair comparison, we trained and tested the reconstruction methods with the same datasets. The performance of the compared methods has cross-validated with three different noise levels, where the standard deviation of noise distribution was set as $\sigma = (10,20,30)$. Later, we summarized the performance of deep models with standard evaluation metrics like PSNR, SSIM, and DeltaE2000.

\subsubsection{Noisy Nona-Bayer Reconstruction}
We performed an extensive evaluation on challenging noisy Nona-Bayer reconstruction by incorporating quantitative and qualitative comparisons.

\begin{table}[!htb]
\begin{center}
\scalebox{.6}{\begin{tabular}{|c|c|c|c|c|c|c|c|c|c|c|}
\hline
\multirow{2}{*}{\textbf{Model}} & \multirow{2}{*}{\textbf{$\sigma$}} & \multicolumn{3}{c|}{\textbf{sRGB Images}}                 & \multicolumn{3}{c|}{\textbf{Linear RGB Images}}        \\ \cline{3-8}  
                       &                        & \textbf{PSNR $\uparrow$}  & \textbf{SSIM $\uparrow$}   & \textbf{DeltaE $\downarrow$}    & \textbf{PSNR $\uparrow$}  & \textbf{SSIM $\uparrow$}   & \textbf{DeltaE $\downarrow$}    \\ \hline \hline
Deepjoint \cite{gharbi2016deep}               & \multirow{5}{*}{\textbf{10}}    & 31.63    & 0.9026   & 3.11     & 39.00     & 0.9464     & 1.64      \\ 
Kokkinos \cite{kokkinos2018deep}                   &                        & 33.08    & 0.9321   & 2.75     & 39.26     & 0.9539     & 1.74      \\ 
DPN \cite{kim2019deep}                    &                        & 33.49    & 0.9390   & 2.62     & 39.84     & 0.9702     & 1.50      \\ 
BJDD \cite{a2021beyond}                   &                        & 34.02    & 0.9440   & 2.56     & 41.40     & 0.9751     & 1.64      \\ 
\textbf{SAGAN (Ours)}                 &                        & \textbf{34.99}    & \textbf{0.9503}   & \textbf{2.18}     & \textbf{43.17}     & \textbf{0.9788}     & \textbf{1.11}      \\ \hline \hline
Deepjoint \cite{gharbi2016deep}                  & \multirow{5}{*}{\textbf{20}}    & 30.22    & 0.8495   & 3.44     & 36.14     & 0.8946     & 1.94      \\ 
Kokkinos \cite{kokkinos2018deep}                   &                        & 31.88    & 0.9080   & 2.97     & 38.18     & 0.9411     & 1.76      \\ 
DPN \cite{kim2019deep}                    &                        & 32.13    & 0.9152   & 2.92     & 38.39     & 0.9572     & 1.68      \\ 
BJDD \cite{a2021beyond}                   &                        & 32.58    & 0.9212   & 2.86     & 39.71     & 0.9619     & 1.86      \\ 
\textbf{SAGAN (Ours)}                   &                        & \textbf{33.33}    & \textbf{0.9290}   & \textbf{2.49}     & \textbf{41.26}     & \textbf{0.9675}     & \textbf{1.32}      \\ \hline \hline
Deepjoint \cite{gharbi2016deep}                  & \multirow{5}{*}{\textbf{30}}    & 28.81    & 0.7913   & 3.90     & 34.05     & 0.8407     & 2.29      \\ 
Kokkinos \cite{kokkinos2018deep}                   &                        & 30.81    & 0.8830   & 3.23     & 36.84     & 0.9203     & 1.95      \\ 
DPN \cite{kim2019deep}                    &                        & 30.96    & 0.8904   & 3.21     & 36.99     & 0.9411     & 1.93      \\ 
BJDD \cite{a2021beyond}                   &                        & 31.42    & 0.8990   & 3.14     & 38.27     & 0.9466     & 2.03      \\ 
\textbf{SAGAN (Ours)}                   &                        & \textbf{32.10}    & \textbf{0.9084}   & \textbf{2.78}     & \textbf{39.59}     & \textbf{0.9525}     & \textbf{1.57}      \\ \hline 
\end{tabular}}
\caption{Quantitative Comparison for Noisy Nona-Bayer reconstruction. }
\label{tab:quanComp}
\end{center}
\end{table}

\vspace{-.5cm}
\textbf{Quantitative Comparison.} Table. \ref{tab:quanComp} demonstrates the performance of the different learning-based methods for Nona-Bayer reconstruction. The proposed SAGAN outperforms the state-of-the-art methods in both sRGB and linear RGB colour spaces. Also, the performance of our SAGAN is consistent among different noise levels. Apart from suppressing noises, our SAGAN can produce more colour-accurate RGB images with dense structural information.

\begin{figure}[!htb]
\begin{center}
\begin{tabular}{cccc}

\multirow{2}{*}{\bmvaHangBox{{\includegraphics[width=3.00cm,height=2.81cm]{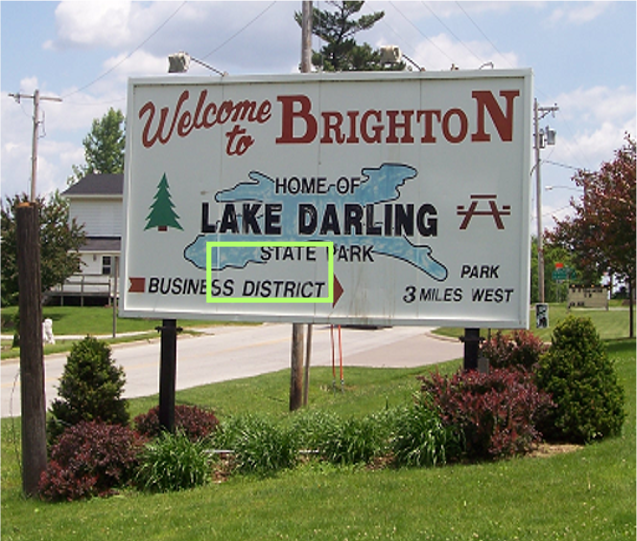}}}}  & 

\hspace{-.45cm}
\bmvaHangBox{{\includegraphics[width=2.0cm, height=1.13cm]{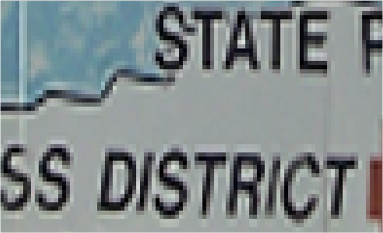}}} & 
\hspace{-.45cm}
\bmvaHangBox{{\includegraphics[width=2.0cm, height=1.13cm]{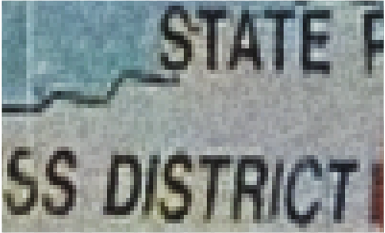}}}& 
\hspace{-.45cm}
\bmvaHangBox{{\includegraphics[width=2.0cm, height=1.13cm]{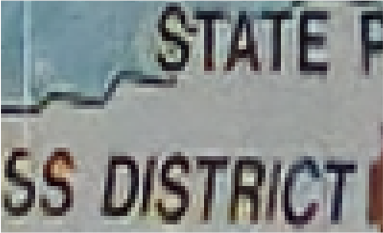}}}\\
&(b)&(c)&(d)\\

& \hspace{-.45cm} \bmvaHangBox{{\includegraphics[width=2.0cm, height=1.13cm]{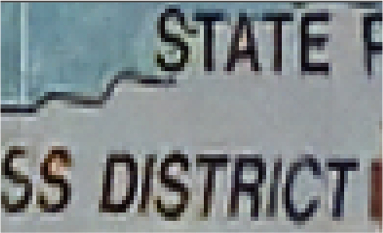}}}& \hspace{-.45cm}
\bmvaHangBox{{\includegraphics[width=2.0cm, height=1.13cm]{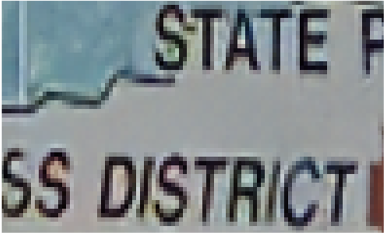}}}& \hspace{-.45cm}
\bmvaHangBox{{\includegraphics[width=2.0cm, height=1.13cm]{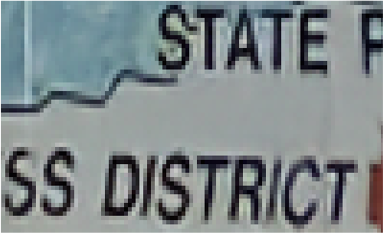}}}\\
(a)&(e)&(f)&(g)

\end{tabular}
\caption{Qualitative Comparison for Noisy Nona-Bayer reconstruction at $\sigma=30$. (a) Ground-truth RGB Image (full). (b) Ground-truth RGB Image (crop). (c) Deepjoint \cite{gharbi2016deep}. (d) Kokkinos \cite{kokkinos2018deep}. (e) DPN \cite{kim2019deep}. (f) BJDD \cite{a2021beyond}. (g) \textbf{SAGAN (Ours)}}
\label{fig:vis}
\end{center}
\vspace{-.6cm}
\end{figure}

\textbf{Qualitative Comparison.} Apart from quantitative comparison, we compared the reconstruction method to visualize their performance. Fig. \ref{fig:vis} illustrates the visual comparison between the existing method and our SAGAN. It can be seen that the proposed SAGAN can reconstruct more natural-looking plausible images with maximum noise suppression. Our novel SAGAN can substantially reduce the visual artefacts that occur due to the non-Bayer CFA pattern. Notably, our proposed adversarial spatial-asymmetric attention strategies allow us to learn perceptually admissible images as similar to the reference images.

\vspace{-.3cm}
\subsubsection{Noisy Bayer Reconstruction}
Typically, Nona-Bayer sensors are capable of forming a Bayer pattern by leveraging the pixel-binning technique. Thus, we have studied our method on noisy Bayer reconstruction to confirm its practicability in real-world scenarios. 

\textbf{Quantitative Comparison.} Table. \ref{tab:quanCompBayer} illustrates the comparison between state-of-the-art methods for noisy Bayer reconstitution on different noise levels. Notably, our SAGAN outperforms the existing methods for noisy Bayer reconstruction as well.

\begin{table}[!htb]
\begin{center}
\scalebox{.6}{\begin{tabular}{|c|c|c|c|c|c|c|c|c|c|c|}
\hline
\multirow{2}{*}{\textbf{Model}} & \multirow{2}{*}{\textbf{$\sigma$}} & \multicolumn{3}{c|}{\textbf{sRGB Images}}                 & \multicolumn{3}{c|}{\textbf{Linear RGB Images}}        \\ \cline{3-8} 
                                &                                 & \textbf{PSNR $\uparrow$}  & \textbf{SSIM $\uparrow$}   & \textbf{DeltaE $\downarrow$}    & \textbf{PSNR $\uparrow$}  & \textbf{SSIM $\uparrow$}   & \textbf{DeltaE $\downarrow$}    \\ \hline \hline
                                
Deepjoint \cite{gharbi2016deep}                        & \multirow{5}{*}{\textbf{10}}    & 33.04          & 0.9262          & 2.80            & 37.89          & 0.9496          & 1.81            \\ 
Kokkinos \cite{kokkinos2018deep}                               &                                 & 34.24          & 0.9412          & 2.664           & 38.45          & 0.9550          & 1.79            \\ 
DPN \cite{kim2019deep}                           &                                 & 36.51          & 0.9593          & 1.88            & 42.80          & 0.9790          & 1.21            \\ 
BJDD \cite{a2021beyond}                             &                                 & 36.68          & 0.9561          & 1.91            & 43.70          & 0.9760          & 1.12            \\ 
\textbf{SAGAN (Ours)}           &                                 & \textbf{37.07} & \textbf{0.9616} & \textbf{1.76}   & \textbf{43.66} & \textbf{0.9756} & \textbf{1.16}   \\ \hline \hline
Deepjoint \cite{gharbi2016deep}                        & \multirow{5}{*}{\textbf{20}}    & 31.08          & 0.8594          & 3.54            & 35.30          & 0.8839          & 2.53            \\ 
Kokkinos \cite{kokkinos2018deep}                               &                                 & 32.37          & 0.9052          & 3.16            & 36.77          & 0.9251          & 2.19            \\ 
DPN \cite{kim2019deep}                           &                                 & 34.22          & 0.9316          & 2.33            & 40.32          & 0.9642          & 1.53            \\ 
BJDD \cite{a2021beyond}                            &                                 & 34.43          & 0.9323          & 2.31            & 41.10          & 0.9596          & 1.42            \\ 
\textbf{SAGAN (Ours)}           &                                 & \textbf{34.56} & \textbf{0.9375} & \textbf{2.20}   & \textbf{42.22} & \textbf{0.9715} & \textbf{1.22}   \\ \hline \hline
Deepjoint \cite{gharbi2016deep}                        & \multirow{5}{*}{\textbf{30}}    & 28.99          & 0.7789          & 4.49            & 32.89          & 0.7997          & 3.38            \\ 
Kokkinos \cite{kokkinos2018deep}                               &                                 & 30.27          & 0.8562          & 3.85            & 34.18          & 0.865           & 2.80            \\ 
DPN \cite{kim2019deep}                           &                                 & 32.32          & 0.8983          & 2.82            & 38.06          & 0.9405          & 1.92            \\ 
BJDD \cite{a2021beyond}                           &                                 & 32.75          & 0.9074          & 2.63            & 38.21          & 0.9298          & 1.75            \\ 
\textbf{SAGAN (Ours)}           &                                 & \textbf{33.28} & \textbf{0.9212} & \textbf{2.41}   & \textbf{40.78} & \textbf{0.9613} & \textbf{1.32}   \\ \hline \hline
\end{tabular}}
\caption{Quantitative comparison for noisy Bayer reconstruction. }
\label{tab:quanCompBayer}
\end{center}
\end{table}

\vspace{-.5cm}

\begin{figure}[!htb]
\begin{center}
\begin{tabular}{cccc}

\multirow{2}{*}{\bmvaHangBox{{\includegraphics[width=3.00cm,height=2.81cm]{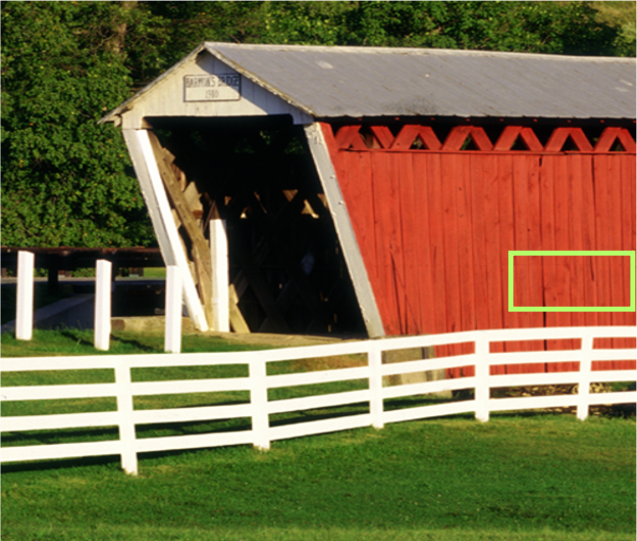}}}}  & 

\hspace{-.45cm}
\bmvaHangBox{{\includegraphics[width=2.0cm, height=1.13cm]{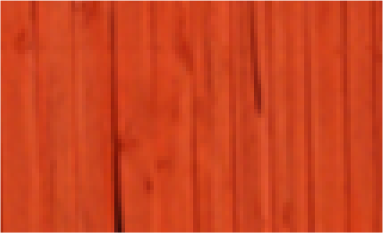}}} & 
\hspace{-.45cm}
\bmvaHangBox{{\includegraphics[width=2.0cm, height=1.13cm]{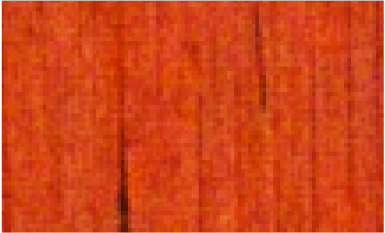}}}& 
\hspace{-.45cm}
\bmvaHangBox{{\includegraphics[width=2.0cm, height=1.13cm]{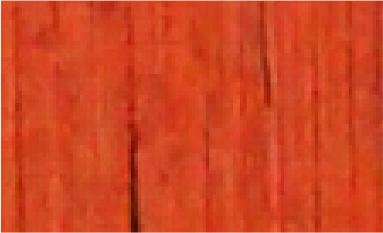}}}\\
&(b)&(c)&(d)\\

& \hspace{-.45cm} \bmvaHangBox{{\includegraphics[width=2.0cm, height=1.13cm]{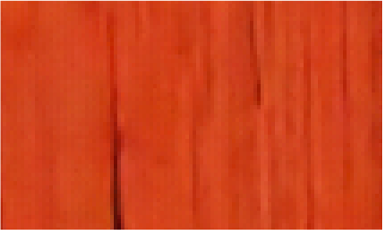}}}& \hspace{-.45cm}
\bmvaHangBox{{\includegraphics[width=2.0cm, height=1.13cm]{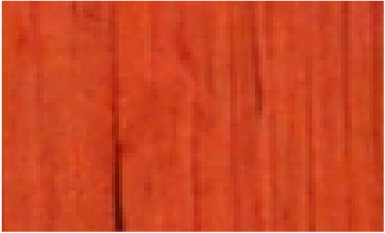}}}& \hspace{-.45cm}
\bmvaHangBox{{\includegraphics[width=2.0cm, height=1.13cm]{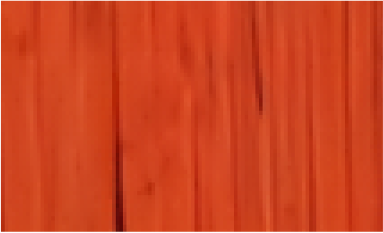}}}\\
(a)&(e)&(f)&(g)

\end{tabular}
\caption{Qualitative Comparison for Noisy Bayer reconstruction at $\sigma=30$. (a) Ground-truth RGB Image (full). (b) Ground-truth RGB Image (crop). (c) Deepjoint \cite{gharbi2016deep}. (d) Kokkinos \cite{kokkinos2018deep}. (e) DPN \cite{kim2019deep}. (f) BJDD \cite{a2021beyond}. (g) \textbf{SAGAN (Ours).}}
\label{fig:visBayer}
\end{center}
\vspace{-.6cm}
\end{figure}

\textbf{Qualitative Comparison.} Fig. \ref{fig:visBayer} visually confirms that the proposed SAGAN can produce plausible images while reconstructing images from noisy Bayer inputs. Also, it can suppress maximum noise by retaining details comparing to its counterparts.

\subsection{Nona-Bayer Reconstruction with Real-world Denoising}

Real-world sensors are typically surrounded by multiple noise sources and can go beyond a synthesized noise. Hence, we studied our method on real-world noisy images also.

\begin{figure}[!htb]
\begin{center}
\begin{tabular}{cccccc}

\hspace{-.45cm}
\bmvaHangBox{{\includegraphics[width=2.1cm,height=2.8cm]{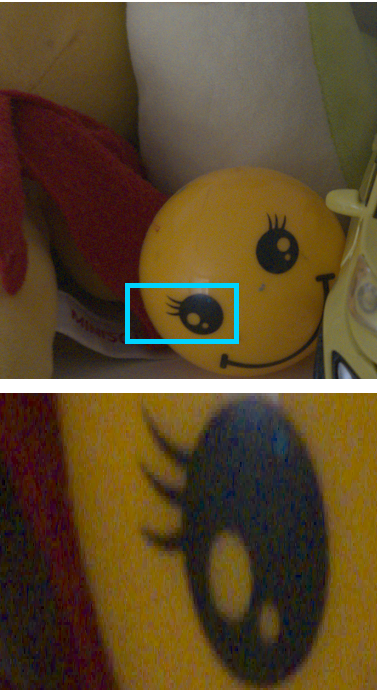}}} & 
\hspace{-.45cm}
\bmvaHangBox{{\includegraphics[width=2.1cm,height=2.8cm]{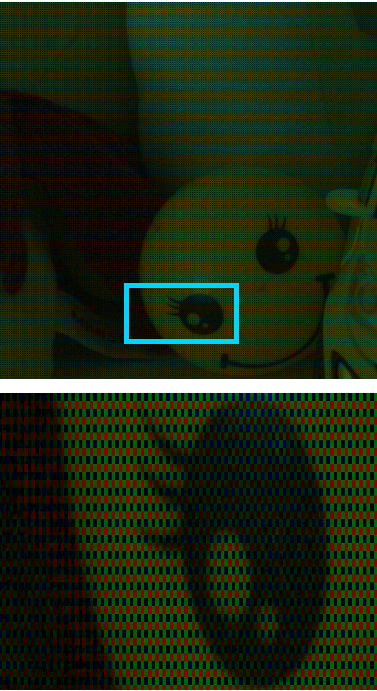}}} & 
\hspace{-.45cm}
\bmvaHangBox{{\includegraphics[width=2.1cm,height=2.8cm]{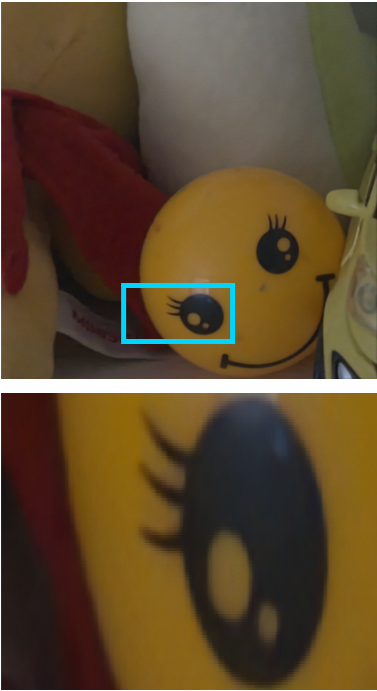}}}& 
\hspace{-.45cm}
\bmvaHangBox{{\includegraphics[width=2.1cm,height=2.8cm]{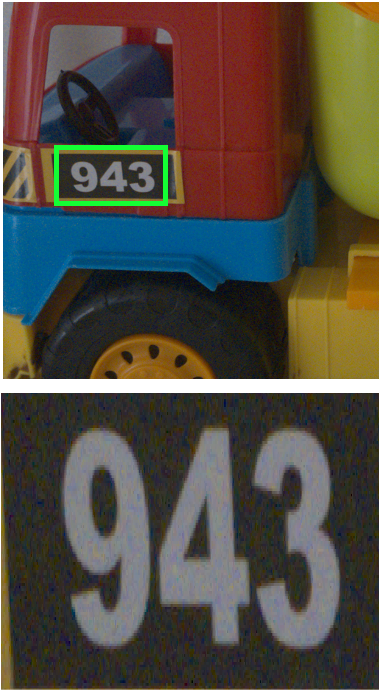}}}& 
\hspace{-.45cm}
\bmvaHangBox{{\includegraphics[width=2.1cm,height=2.8cm]{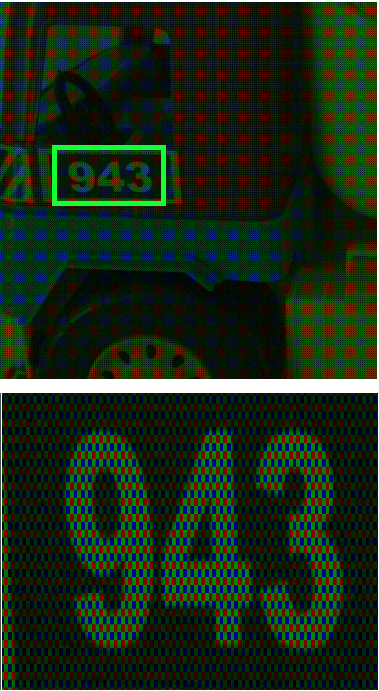}}}& 
\hspace{-.45cm}
\bmvaHangBox{{\includegraphics[width=2.1cm,height=2.8cm]{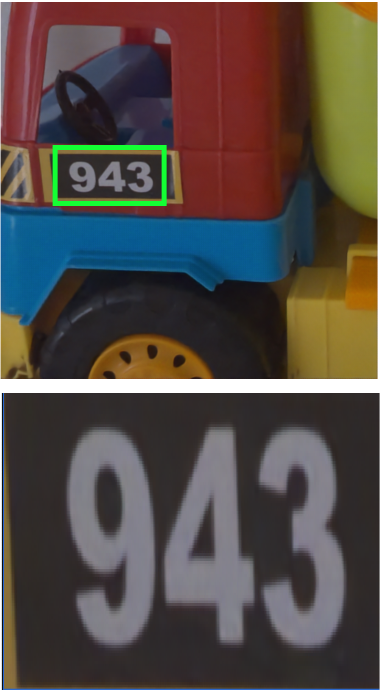}}}\\
(a)&(b)&(c)&(d)&(e)&(f)\\

\end{tabular}
\caption{Nona-Bayer reconstruction with real-world noise suppression. (a) \& (d) Noisy image (RAW). (b) \& (e) Noisy Nona-Bayer (input). (c) \& (f) \textbf{Reconstructed with SAGAN}.}
\label{fig:realRecon}
\end{center}
\end{figure}

\textbf{Visual Results.} Fig. \ref{fig:realRecon} depicts the performance of proposed SAGAN on real-world denoising with Nona-Bayer reconstruction. It can be seen that our method can handle real-world noise and can produce visually plausible images without any visual artefacts.

\begin{wraptable}{!htb}{0.36\textwidth}
\vspace{-.7cm}
\begin{center}
\scalebox{.65}{\begin{tabular}{|c|c|c|}
\hline
\textbf{Method} & \textbf{RAW (Visualised)} & \textbf{SAGAN (Ours)} \\ \hline \hline
\textbf{MOS $\uparrow$ }    &         0.13                   &     \textbf{0.87}    \\ \hline
\end{tabular}}
\caption{User study for real-world noisy Nona-Bayer reconstruction. }
\label{fig:tabMOS}
\end{center}
\vspace{-.5cm}
\end{wraptable}
\vspace{-.5cm}

\textbf{User Study.}
We performed a blindfold user study to verify the practicability of our proposed method. Therefore, we showed image pairs comprising our reconstructed and noisy (RAW) image to the random users and asked them to select their preferred image from each image pair. Later, we calculated the mean opinion score (MOS) to summarized user preferences. Our proposed method can substantially score higher MOS, as shown in Table. \ref{fig:tabMOS}.

\vspace{-.2cm}
\begin{table}[!ht]
\begin{center}
    
\scalebox{.65}{\begin{tabular}{|c|c|c|c|c|c|c|c|c|c|c|}
\hline
\multirow{2}{*}{\textbf{Model}} & \multirow{2}{*}{\textbf{Base}} & \multirow{2}{*}{\textbf{SA}} & \multirow{2}{*}{\textbf{PCL}} & \multirow{2}{*}{\textbf{GAN}} & \multicolumn{3}{c|}{\textbf{sRGB Images}}       & \multicolumn{3}{c|}{\textbf{Linear RGB Images}} \\ \cline{6-11} 
                                &                                &                              &                               &                               & \textbf{PSNR $\uparrow$}  & \textbf{SSIM $\uparrow$}   & \textbf{DeltaE $\downarrow$}    & \textbf{PSNR $\uparrow$}  & \textbf{SSIM $\uparrow$}   & \textbf{DeltaE $\downarrow$}  \\ \hline \hline
BaseNet                         &      \cmark                           &            \xmark                   &      \xmark                           &          \xmark                       & 22.26          & 0.5800            & 10.63           & 24.65          & 0.6112          & 9.42                \\ 
BaseGAN                         &      \cmark                          &                  \xmark              &   \xmark                              &           \cmark                    &  27.45             &  	0.7730	        &       6.09          &      26.52	       &    0.6684            &  	8.48              \\ 
SANWP                            &       \cmark                         &                   \cmark           &      \xmark                          &              \xmark                  &    31.99	           &       0.9125	       &     2.89            &      35.25	         &     0.8191          &        1.87         \\ 
SAN                             &        \cmark                         &               \cmark                &     \cmark                           &              \xmark                 & 32.75          & 0.9240          & 2.58            & 36.59          & 0.9588          & 1.42             \\ 
\textbf{SAGAN}                  &      \cmark                          &               \cmark               &     \cmark                          &            \cmark                   & \textbf{33.47} & \textbf{0.9292} & \textbf{2.48}   & \textbf{41.34} & \textbf{0.9663} & \textbf{1.33}               \\ \hline
\end{tabular}}
\caption{ Quantitative evaluation on proposed SAGAN. It can be seen that our proposed components have a meaningful impact on the noisy Nona-Bayer reconstruction.}
\label{fig:tabAbl}
\end{center}
\end{table}

\vspace{-1.0cm}
\subsection{Ablation Study}

The practicability of our proposed spatial-asymmetric attention with adversarial training has been verified with sophisticated experiments. We removed our proposed components like spatial-asymmetric attention block, PCL, and SAGAN discriminator from the network architectures. Later, we incorporated each of them individually and summarized the practicability of these components with quantitative and qualitative evaluation. 

\textbf{Quantitative Evaluation.} Table. \ref{fig:tabAbl} illustrates the practicability of our proposed spatial-asymmetric attention and adversarial guidance in both colour spaces.  For simplicity, we depicted the mean performance of each model on different noise levels (i.e., $\sigma = 10, 20, 30$). It is visible that our proposed components play a substantial role in Nona-Bayer reconstruction.

\begin{figure}[!htb]
\begin{center}
\begin{tabular}{cccccc}

\hspace{-.45cm}
\bmvaHangBox{{\includegraphics[width=2.1cm,height=2.8cm]{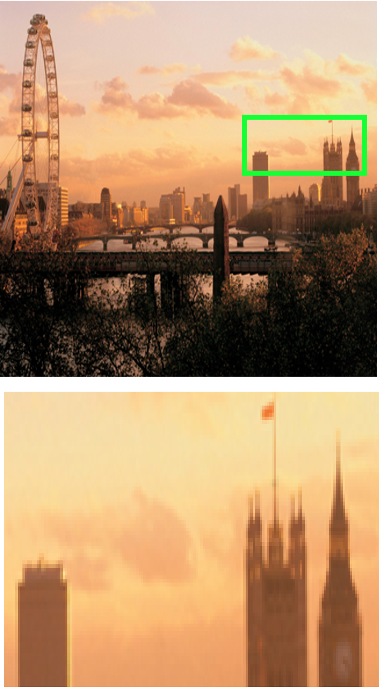}}} & 
\hspace{-.45cm}
\bmvaHangBox{{\includegraphics[width=2.1cm,height=2.8cm]{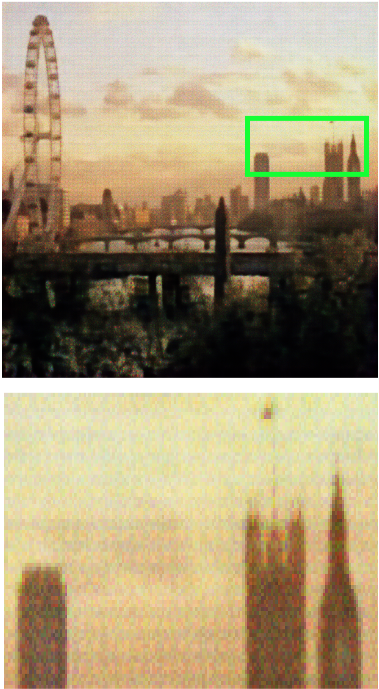}}} & 
\hspace{-.45cm}
\bmvaHangBox{{\includegraphics[width=2.1cm,height=2.8cm]{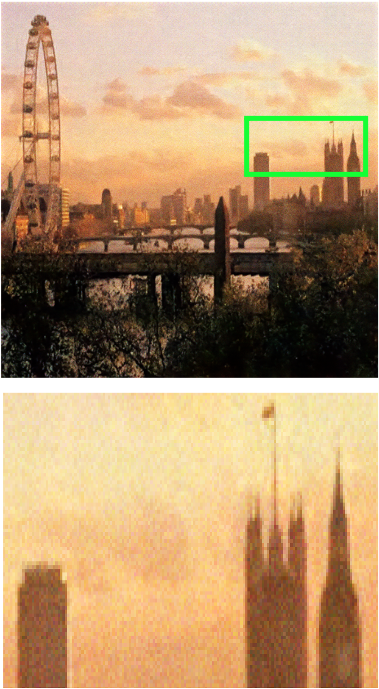}}}& 
\hspace{-.45cm}
\bmvaHangBox{{\includegraphics[width=2.1cm,height=2.8cm]{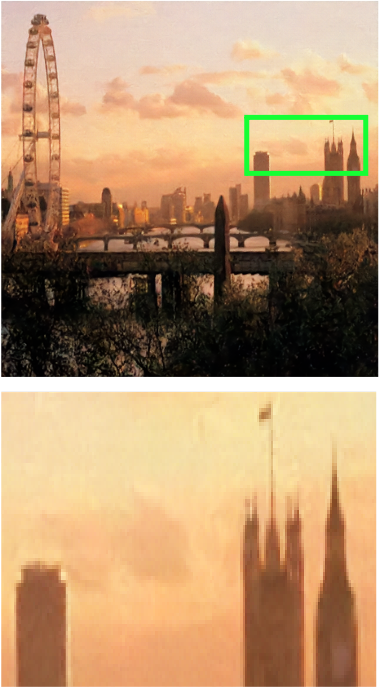}}}& 
\hspace{-.45cm}
\bmvaHangBox{{\includegraphics[width=2.1cm,height=2.8cm]{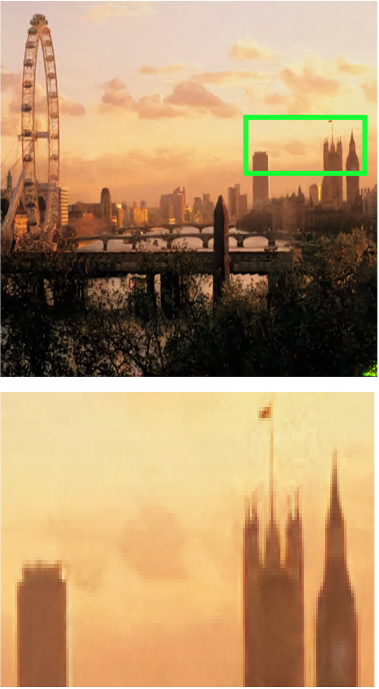}}}& 
\hspace{-.45cm}
\bmvaHangBox{{\includegraphics[width=2.1cm,height=2.8cm]{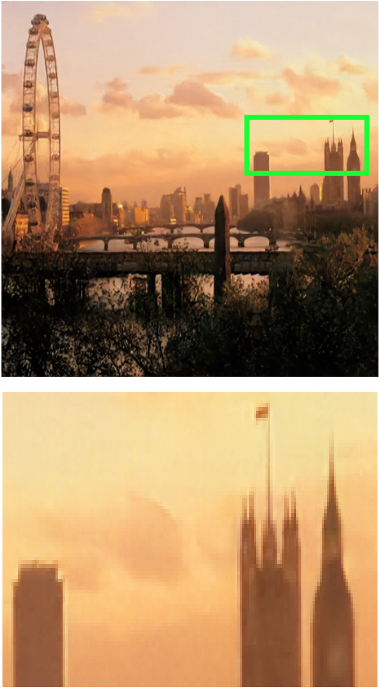}}}\\
(a)&(b)&(c)&(d)&(e)&(f)\\

\end{tabular}
\caption{Qualitative evaluation of our proposed SAGAN at $\sigma=30$. Our proposed component can substantially reduce visual artefacts and produce perceptually plausible images (best viewed in colour and zoomed). (a) Ground-truth RGB Image. (b) BaseNet. (c) BaseGAN. (d) SANWP. (e) SAN. (f) SAGAN.}
\label{fig:visAbl}
\end{center}

\vspace{-.5cm}
\end{figure}

\textbf{Qualitative Evaluation.} Fig. \ref{fig:visAbl} illustrates the visual comparison between SAGAN variants. Also, it confirms that our proposed spatial-asymmetric attention can substantially reduce the visual artefacts, while our adversarial training helps us recover texture with natural colours. Additionally, PCL has helped us to maintain a colour consistency across different colour spaces.

\vspace{-.5cm}
\subsection{Discussion}
Proposed SAGAN comprises a total of 29,448,766 trainable parameters. As being a fully convolutional network architecture, the proposed network can be inference with different resolution images. In our setup, we found it takes only 0.80 sec in reconstructing a $1024 \times 1024 \times 3$. It is noteworthy that our proposed SAGAN does not incorporate any pre/post-operations. Therefore, the inference time with similar hardware is expected to remain constant. Despite showing an admissible inference time on a desktop environment, we failed to study SAGAN on a real-world mobile setup due to hardware limitations. Nevertheless, the proposed SAGAN reveals a promising aspect of noisy Nona-Bayer reconstruction through deep learning. Please see the supplementary material for implementation details and more results of our proposed SAGAN. 

\vspace{-.2cm}
\section{Conclusion}
We proposed a novel end-to-end deep method for reconstructing  RGB images from a challenging Nona-Bayer CFA. Notably, our proposed method incorporates a novel spatial-asymmetric attention mechanism with adversarial training. We studied the feasibility of our SAGAN on different colour spaces and diverse data samples. Experimental results illustrate that our SAGAN can outperform the existing methods in both quantitative and qualitative comparisons. However, due to hardware constraints, we failed to evaluate the performance of our SAGAN by deploying it into real mobile hardware. It has planned to study the practicability of a deep method like SAGAN for reconstructing images from Nona-Bayer along with Quad-Bayer CFA on real mobile hardware in the foreseeable future.

\end{document}